\documentclass{article}
\usepackage[english]{babel}
\pdfoutput=1
\usepackage[sort,compress]{cite}
\usepackage{url}
\usepackage{subfig}
\usepackage{multirow}
\usepackage{algorithm}
\usepackage[noend]{algpseudocode}
\usepackage{amsmath}
\usepackage{xcolor}
\usepackage{authblk}
\usepackage{graphicx}
\usepackage{multirow}
\usepackage{amssymb}
\makeatletter
\def\BState{\State\hskip-\ALG@thistlm}
\makeatother

\begin{document}

\title{Audio Captioning with Composition of Acoustic and Semantic Information}

\author[ ]{Ayşegül Özkaya Eren}
\author[ ]{Mustafa Sert}
\affil[ ]{Department of Computer Engineering\\
	Başkent University\\ 
	Ankara, Turkey}

\affil[ ]{\textit{21610279@mail.baskent.edu.tr}}
\affil[ ]{\textit{msert@baskent.edu.tr}}
\date{}
\maketitle

\begin{abstract}
	
Generating audio captions is a new research area that combines audio and natural language processing to create meaningful textual descriptions for audio clips. To address this problem, previous studies mostly use the encoder-decoder based models without considering semantic information. To fill this gap, we present a novel encoder-decoder architecture using bi-directional Gated Recurrent Units (BiGRU) with audio and semantic embeddings. We extract semantic embedding by obtaining subjects and verbs from the audio clip captions and combine these embedding with audio embedding to feed the BiGRU-based encoder-decoder model. To enable semantic embeddings for the test audios, we introduce a Multilayer Perceptron classifier to predict the semantic embeddings of those clips. We also present exhaustive experiments to show the efficiency of different features and datasets for our proposed model the audio captioning task. To extract audio features, we use the log Mel energy features, VGGish embeddings, and a pretrained audio neural network (PANN) embeddings. Extensive experiments on two audio captioning datasets Clotho and AudioCaps show that our proposed model outperforms state-of-the-art audio captioning models across different evaluation metrics and using the semantic information improves the captioning performance.


\end{abstract}

\section{Introduction}
Audio captioning is a newly proposed task to describe the content of an audio clip using natural language sentences \cite{DBLP:journals/corr/DrossosAV17}. The purpose of creating captions is not only finding the objects, events, or scenes in the given audio clip but also finding relations between them and generating meaningful sentences. It has great potential for real-life applications such as assisting hearing impaired people and understanding environmental sounds. Additionally, since smart audio-based and video surveillance systems use audio signals, audio signal analysis is a critical research area for surveillance systems. These systems can be used for recognizing activities, detecting events, anomalies and finding semantic relations between video and audio for child-care centers, nursing homes, smart cities, elevators, etc. \cite{DBLP:journals/corr/CroccoCTM14,10.1145/3322240,8633626}.

\begin{figure*}[t]
	\centering
	
	\subfloat[]{%
		\raisebox{-0.5\height}{\includegraphics[width=.4\linewidth]{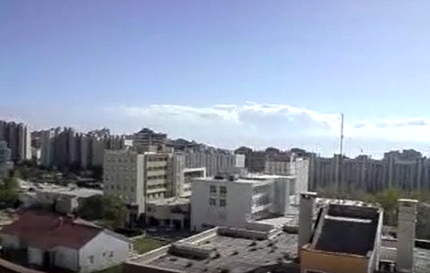}}}
	\hspace{0.1\textwidth}
	\subfloat[]{%
		\raisebox{-0.5\height}{\includegraphics[width=.4\linewidth]{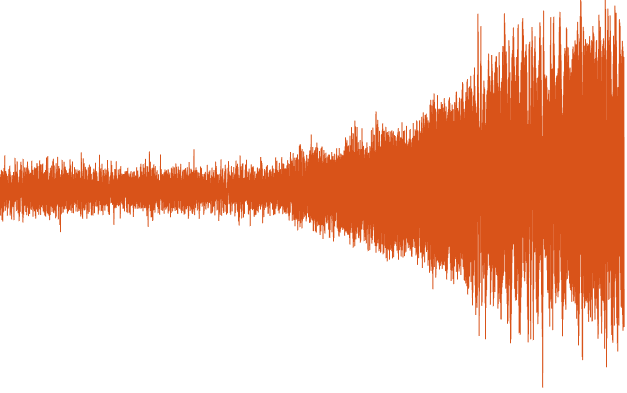}}}
	\caption{A sample scene (a) from audio-enabled video surveillance [5] and its partial sound wave (b). Without audio, there are only buildings on the scene. We can also capture the siren sound by using
		audio modality.}
\end{figure*}

In the field of audio signal processing, a number of tasks, such as audio event classification/detection \cite{6287923}, acoustic scene recognition \cite{7760424,8959049}, and audio tagging \cite{DBLP:journals/corr/KongXWP16} have received much attention over the past few years. In the audio event detection task, the main aim is to identify (overlapping) sound events occurring in the audio clip along with their starting and ending times. The audio tagging task assigns predefined labels to a given audio segment, whereas the acoustic scene recognition task concerns the understanding of the acoustics of the environment and assigns labels to it. However, audio captioning is quite a higher level of abstraction of these tasks in the sense of generating descriptive sentences in a natural language. In audio-enabled video surveillance systems, these sentences can be used for the understanding of video scenes and possible abnormality detection within them, as well as indexing and retrieval of video (Figure 1). 

Captioning is firstly used for describing images and numerous studies have been conducted \cite{DBLP:journals/corr/XuBKCCSZB15,DBLP:journals/corr/ChoCB15}. This is followed by the video captioning task, which aims to generate captions for video clips \cite{8356255,DBLP:journals/corr/abs-1804-00819}. Audio captioning task is first described in \cite{DBLP:journals/corr/DrossosAV17}. Drossos et al. propose an encoder-decoder model with three Bi-directional Gated Recurrent Unit (BiGRU) layers in the encoder and two Gated Recurrent Unit (GRU) layers in the decoder to generate audio captions by means of an attention mechanism. They use the log Mel energies as audio features and a commercial dataset ProSound Effects \cite{prosound} in their experiments. ProSound Effects dataset includes a set of keywords as audio captions. Although this study achieves generating audio captions from the audio clips, the results are not proper sentences. Wu et al.\cite{DBLP:journals/corr/abs-1902-09254} present another attempt in the field of audio captioning. Their model is an encoder-decoder model with one GRU layer in the encoder and one GRU layer in the decoder. Also, they introduce a new audio captioning dataset for the Chinese language. This model can produce audio captions but the model tends to produce repetitive sentences. An encoder-decoder model with semantic attention for generating captions for audios in the wild is presented by Kim et al. to produce semantically meaningful audio captions and to improve audio captioning performance. They contribute a large-scale dataset AudioCaps of 46K audio clips \cite{kim-etal-2019-audiocaps}. For semantic attention, they extract the words from the captions and apply the nearest neighbor approach to the AudioSet\cite{7952261} dataset to retrieve the nearest labels as attribute words. These attributes are added to the model as semantic information. Drossos et al. introduce a publicly available audio captioning dataset called Clotho \cite{Drossos_2020} and present the results with the method in \cite{DBLP:journals/corr/DrossosAV17}. Çakır et. al presents new results using the Clotho dataset on the audio captioning task \cite{akr2020multitask}. They proposed a model to capture the words that are used infrequently but informative. A multi-task regularization method is applied to solve the distribution of words problem in the audio captions. Nguyen et al. propose another model using a temporal sub-sampling to the audio input sequence \cite{nguyen2020temporal}. Bi-directional RNN-based encoder is used as the model architecture. They present their results on the Clotho dataset.

Semantic information is previously studied to improve the performance of the image and video captioning tasks \cite{XIAO2019285,8499357}. In the audio captioning task, the semantic attributes are firstly used in \cite{kim-etal-2019-audiocaps}. They use AudioSet \cite{7952261} labels as semantic attributes by using the labels of the nearest video clip.

Since most existing approaches in the audio captioning task use the encoder-decoder model without semantic concepts, our motivation is to improve audio captioning recognition performance by using semantic information along with the audio embeddings. To address this problem, we add semantic concepts using subject-verb embeddings. We propose a novel model using the Pretrained Audio Neural Networks (PANNs) \cite{kong2019panns} as a feature extractor for audio feature embedding and Word2Vec \cite{DBLP:journals/corr/MikolovSCCD13} for word embedding since their performance is shown in audio classification tasks \cite{DBLP:journals/corr/abs-1905-01926}. The preliminary results of this study has been published in [25]. The extensions and the core contributions of our study are as follows:
\begin{itemize}
	\item We present a novel encoder-decoder model, namely RNN-GRU-EncDec for the audio captioning task. Audio and semantic embeddings are extracted and added to the proposed model to improve captioning performance.
	\item Different from the previous studies in the audio captioning tasks, we extract subjects and verbs from the captions in the datasets to obtain semantic embeddings.
	\item We present exhaustive experiments to show the contribution of different audio features such as the log Mel energies, VGGish embeddings \cite{7952132}, and PANNs embeddings. To the best of our knowledge, this is the one of the earliest papers \cite{9327916} employing the PANNs as a feature extractor in the audio captioning field.
	\item We demonstrate our results on two new audio-captioning datasets to validate the effectiveness of our proposed model.
\end{itemize}
The organization of the paper is as follows. Section 2 introduces our proposed method. We present our experimental results and evaluations in Section 3. Finally, we give concluding remarks and possible future directions in Section 4.

\section{Proposed Method}
Our main aim is to generate meaningful captions for a given audio clip. Mathematically: 
\begin{equation}
\begin{aligned}
\theta^{\star}=\underset{\theta}{\operatorname{argmax}} 
\sum_{A,\textbf{c}} \log p(\textbf{c}|A;\theta) 
\end{aligned}
\end{equation} 
We aim to maximize the probability of the caption $\textbf{c}$ for a given audio clip $A$ according to model parameters $\theta$. Since captions are vectors of words, $\textbf{c}$ refers to the caption of a given audio clip.

\begin{equation}
\begin{aligned}
\log p(\textbf{c}|A) = \sum_{t=0}^{N} \log p(c_t|A,c_0,...,c_{t-1}) 
\end{aligned}
\end{equation} 
where, $N$ is the number of words and $c_0$ to $c_{t-1}$ is the words in the given caption.

The overall structure of our proposed model is given in Figure 2. The overall architecture consists of the following modules: The audio embedding extractor, subject-verb embedding extractor, and sequence modeling which is based on RNN-GRU encoder-decoder (RNN-GRU-EncDec). 

\begin{figure*}[t]
	\centering	
	\subfloat[]{\includegraphics[width=1.0\linewidth]{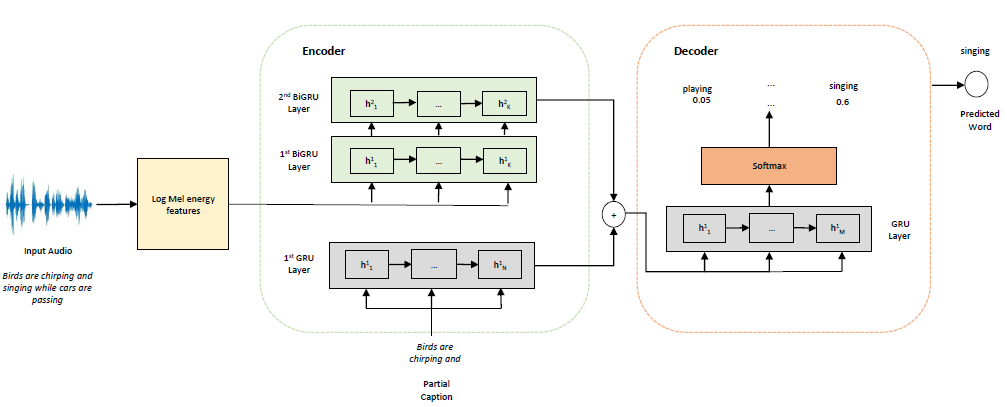}\label{fig:f5}}
	\hspace{1cm}
	\subfloat[]{\includegraphics[width=1.0\linewidth]{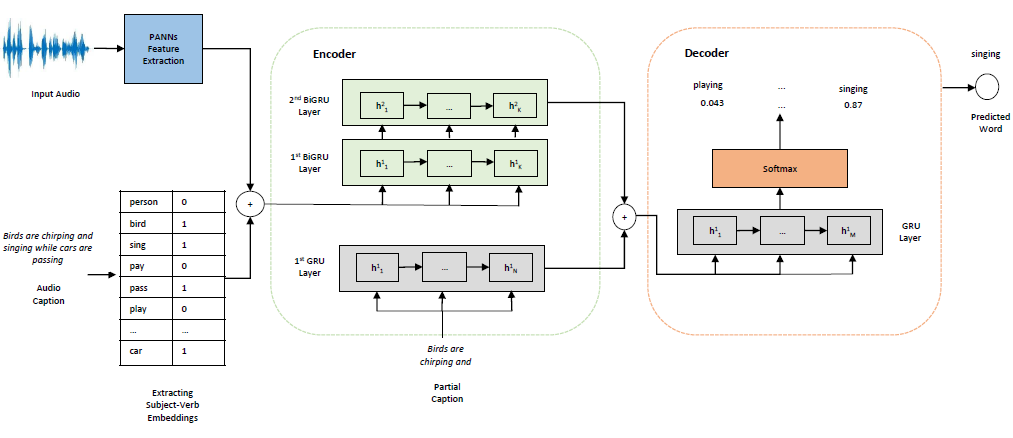}\label{fig:f6}}
	\hspace{1cm}
	\vspace*{10pt}
	\caption{The architecture of the proposed encoder-decoder model RNN-GRU-EncDec with the log Mel energy features (a) and the architecture of the proposed model with PANNs features and subject-verb embeddings (b).}
\end{figure*}

In RNN-GRU-EncDec, first the log Mel energy features are extracted. In the encoder phase, the log Mel energy features and partial captions are encoded separately. In the decoder phase, the output of the encoder is decoded in a GRU layer. The decoder outputs a probability distribution of the unique words in the related dataset. The word that has maximum probability is selected as the predicted word and added to the partially predicted caption until the $<$eos$>$ token is captured as the predicted word. 

In the proposed model with PANNs features and subject-verb embeddings, we first extract PANNs audio embeddings from each audio signal. Afterward, subject-verb embeddings are extracted from the captions for each audio clip. Then audio embeddings and subject-verb embeddings are concatenated. In the encoder phase, these concatenated embeddings and partial captions are encoded separately. In the decoder phase, the output of the encoder is decoded in a GRU layer. The decoder outputs a probability distribution of the unique words in the related dataset. The word that has maximum probability is selected as the predicted word and added to the partially predicted caption until the \texttt{$<$eos$>$} token is captured as the predicted word. The details of these modules are described in Section 2.1, 2.2, 2.3, and 2.4.
\subsection{Audio Features}
\subsubsection{Acoustic Content}

	We extract the log Mel energy features using 96 ms Hamming window with 50\% overlap and obtain 64 log Mel energies for each frame similar to \cite{Drossos_2020}. We set the frequency band to 125-7500 Hz. The log Mel energy features denoted as  $\textbf{x}=[x_1,...,x_T], x_t \in \mathbb{R}^{64}$ , where $x_t$ is a vector that contains 64 features of the audio clip and $T$ is the number of audio frames. 
\subsubsection{Audio Embeddings}
We use the VGGish model and PANNs embeddings to extract audio features. 

VGGish model is pre-trained on the AudioSet \cite{7952261}. The AudioSet is a large-scale audio event dataset and contains 2,084,320 human-labeled 10-second sound clips representing 632 audio event classes. The video clips are in different lengths but the labels represent a 10-second interval of the video clips.

Previous studies show that VGGish embeddings achieve good results compared with hand-crafted audio features in audio classification tasks \cite{DBLP:journals/corr/abs-1905-01926, Basbug2019}. In order to extract audio embedding, we first extract the log Mel spectrograms from audio clips. The length of the clips varies between 15 to 30 seconds. Since the length of the longest audio record is 30 seconds, we apply zero-padding to the audio records which are shorter than 30 seconds. We resample them to 16 Khz. We choose window-size of 96 milliseconds (ms) with 50\% overlap. We set the number of Mel filters to 64 similar to \cite{Drossos_2020} and frequency band to 125-7500 Hz. VGGish model extracts 128-dimensional feature vector for each second. After applying VGGish model, we obtain audio features denoted as $\textbf{x}=[x_1,...,x_T], x_t \in \mathbb{R}^{128}$ , where $x_t$ is a vector that contains 128 features of the audio clip and $T$ is the number of audio frames according to 96 ms window-sizes and 50\% overlaps.

Similarly, the PANNs are pre-trained on the AudioSet \cite{7952261}. The PANNs explore the presence probability of the AudioSet sound classes for the given audio record. The PANNs extract audio features denoted as $\textbf{x}=[x_1,...,x_T], T \in \mathbb{R}^{2048}$, where 2048 is the feature for an audio clip. Among the different PANNS architectures, we use Wavegram-Logmel-CNN14 model as a feature extractor.

\subsection{Subject-Verb Embedding}

The subjects and verbs are informative entities within a sentence and we believe that using those entities as semantic embedding is important to better capture the content of a sentence. To form those embeddings, we use audio captions in the training datasets.

For extracting semantic embeddings, the subject-verb embedding vectors are obtained separately for each dataset. First, each audio caption of each audio record is processed by Stanford Parser and the subject and verbs of the captions are extracted. To reduce the dimension, we use the root forms of the subjects and verbs. Then, subjects and verbs are collected by eliminating repeated words and the subject-verb embedding list is created. The algorithm for extracting subject-verb embeddings is given in Algorithm 1.
\begin{algorithm}
	\caption{Extracting Subject-Verb Embedding}\label{sve}
	\textbf{Input}: Sets of $\textbf{c}_j$ $\in$ \textbf{C}, where \textbf{C} refers to the Caption List in given dataset, \\
	 $\textbf{c}_j$ refers to the caption of given audio in Caption List
	
	\textbf{Output}: Subject-Verb Embedding (SVE) of the dataset
	
	\begin{algorithmic}[1]
		\State \texttt{SVE $\leftarrow$ $\emptyset$;}
		\State \texttt{subjectVerbCorpus $\leftarrow$ $\emptyset$;}
		\State \texttt{$J$ $\leftarrow$ Number of Captions in Caption List}
		
		\For{\texttt{$j$=1,...,$J$}}
		\State \texttt{Get subjects of $\textbf{c}_j$ }
			\If {subjectVerbCorpus does not contain subjects of $\textbf{c}_j$}
			 \State add subjects of $\textbf{c}_j$ to subjectVerbCorpus	\EndIf 
	
		\State \texttt{Get verbs of $\textbf{c}_j$ }
			\If {subjectVerbCorpus does not contain verbs of $\textbf{c}_j$}
			 \State add verbs of $\textbf{c}_j$ to subjectVerbCorpus	\EndIf 
		\EndFor		
	
     	\State \texttt{$K$ $\leftarrow$ subjectVerbCorpus.size}
	
		\For{\texttt{$j$=1,...,$J$}}
		\For{\texttt{$k$=1,...,$K$}}
		\If {$\textbf{c}_j$ contains \texttt{subjectVerbCorpus[k]}}
		\texttt{SVE[j][k]=1}
	    \Else{}
	   	\texttt{SVE[j][k]=0}	   	     
		\EndIf		
		\EndFor
		\EndFor
		
	\end{algorithmic}
\end{algorithm}

Let $\textbf{$y_j$}=[y_{j1},...,y_{jK}] \in \mathbb\{$0,1$\}^{K}$ is the subject-verb vector of the related dataset where $K$ depends on the size of the subject-verb embeddings, $j$ is the $j^{th}$ audio clip. The subject-verb embedding vector is calculated for each audio clip in the training dataset. If $j^{th}$ audio clip contains $y_{jk}$, then $y_{jk}$=0, otherwise $y_{jk}$=1.

During the test phase, we need to predict the subject-verb embeddings. This multilabel classification task stage is conducted through a multilayer perceptron (MLP). In this stage, firstly the PANNs features of the test audio clips are extracted. We designed a six hidden layered architecture, empirically. Batch-size is chosen as 64 and the dropout rate is chosen as 0.5 for input connections, experimentally. ReLU activation function is used for every hidden layer.

For each test audio clip, the subject-verb embedding vector $\textbf{$y_j$}=[y_{j1},...,y_{jK}]$ is predicted using MLP. Let $\textbf{$\overline{y}_j$}=[\overline{y}_{j1},...,\overline{y}_{jK}]$ be probabilities of each subject-verb set for $j^{th}$ test audio clip. We find $\textbf{$\overline{y}_j$} = MLP(x_j)$ where $\textbf{$x_j$}$ represents the audio features of $j^{th}$ audio clip. The subject-verb embeddings extraction architecture of the test audio clips is given in Figure 3. We trained the MLP on the training split of the corresponding dataset for 100 epochs. The minimum validation error is obtained in the $90^{th}$ epoch for the MLP model. The loss and validation loss graph is presented in Figure 4. 

\begin{figure}[t]
	\centering
	\includegraphics[width=1.0\linewidth]{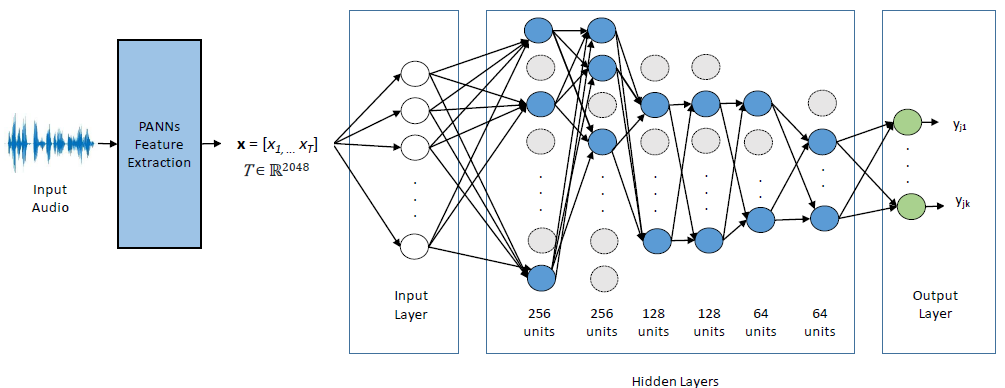}
	\caption{Extracting Subject-Verb Embeddings for the test audio clips. Firstly, PANNs features of the test audio records are extracted. Then, the previously trained MLP model is used to predict subject-verb embeddings. Blue cells present the active neurons and grey cells present the randomly omitted neurons (dropout).}
	\label{figMLP}
\end{figure}
\begin{figure}[t]
	\centering
	\includegraphics[width=0.6\linewidth]{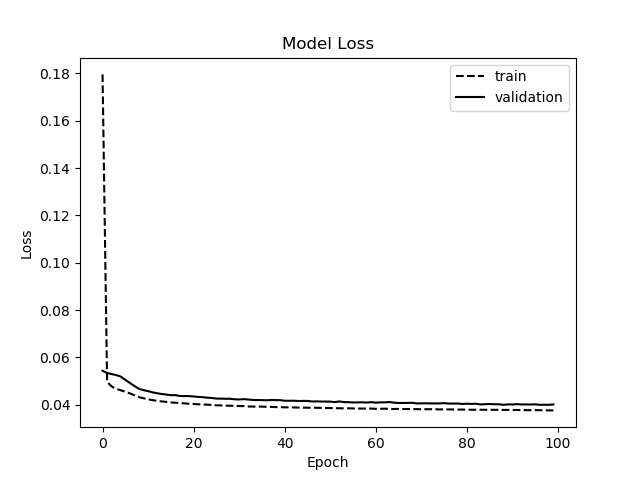}
	\caption{The train-validation loss of the MLP model}
\end{figure}

Finally, the subject-verb embedding vector and audio feature embeddings are concatenated for each audio clip to feed the encoder.

\subsection{Encoder}

The encoder model takes three inputs for the encoding stage which are audio embeddings, subject-verb embeddings, and partial captions. Encoding audio and subject-verb embeddings are the first part of the encoder. We concatenate audio embeddings and subject-verb embeddings before the encoding stage.

We use GRU to learn dependencies between audio frames and subject-verb embeddings in a given audio clip and sequences of words in captions since it reduces the number of parameters in the model \cite{DBLP:journals/corr/Lipton15}. The GRU reads the whole sequence and produces one output. The formulation of a GRU cell in our model is given as:
\begin{equation}
\begin{aligned}
z_t = \sigma(W_z([h_{t-1},x_t]))
\end{aligned}
\end{equation} 
\begin{equation}
\begin{aligned}
r_t = \sigma(W_r([h_{t-1},x_t]))
\end{aligned}
\end{equation} 
\begin{equation}
\begin{aligned}
\hat{h}_t = tanh(W([r_t * h_{t-1},x_t]))
\end{aligned}
\end{equation} 
\begin{equation}
\begin{aligned}
h_t = (1-z_t) * h_{t-1} + z_t * \hat{h}_t
\end{aligned}
\end{equation} 
where  $z_t$ is the update gate at time step t, $x_t$ is the input for time step t. $W$ represents the weights, $\sigma$ is the sigmoid function, and $h_t$ is the hidden state in time step t.

In order to obtain text embeddings, we extract word embedding using the Word2Vec model due to its superiority compared with the one-hot-encoding representation\cite{DBLP:journals/corr/MikolovSCCD13}. We train the Word2Vec model using the captions in the training/development split of the utilized dataset. As a result, we generate $E=[e_1,...,e_i]$  to represent each word vector in the dataset vocabulary, where $e_i\in \mathbb{R}^{256}$ and 256 is the feature dimension of word embedding of each word. We use these pre-trained embeddings to initialize weights in the embedding layer of our model. It is not used in the testing phase.

Unlike feed-forward GRU, BiGRU is able to capture information not only from the past and the current state but the sequence is also reversed in time. Since an audio signal is composed as temporal sequences of frames, we use BiGRU to learn the relationship between audio time steps. We use two BiGRU layers in our design. In the encoding stage of our model, the first BiGRU layer has 32 cells and the second has 64 cells, which are selected empirically. For text encoding, Word2Vec model weights are used to initialize our model’s word embedding layer. This embedding is given to the first GRU layer which has 128 cells. This GRU is used to learn word sequences. In order to combine encoded audio and text, we use the concatenation method.

\subsection{Decoder}

The decoder predicts the partial caption word by word using encoded audio, subject-verb embeddings, and previous partial captions. GRU is used to decode audio and text representations. The proposed GRU layer consists of 128 cells. We use the $Softmax$ after the fully connected layer. The decoder performs the prediction word by word and a sequence of the predicted words gives the caption. Our proposed RNN-GRU-EncDec training algorithm is given in Algorithm 2.

\begin{algorithm}[ht]
	\caption{Training process of RNN-GRU-EncDec}\label{audioCaptioning}
	\textbf{Input}: Sets of $\textbf{x}_j$ $\in$ \textbf{A}, where \textbf{A} refers to the audio features in given dataset,  $\textbf{x}_j$ refers to the features of $j^{th}$ audio clip.
	Sets of $\textbf{c}_j$ $\in$ \textbf{C}, where \textbf{C} refers to the Caption List in given dataset,  $\textbf{c}_j$ refers to the caption of $j^{th}$ audio clip in Caption List. $numEpoch$ number of epoches. 
	$batchSize$ number of batch size. $w_1$ to $w_{t-1}$ are the partial caption words and $w_t$ is the target word based on partial caption(previous) words.
	
	\textbf{Output}: $w_t$ is the target word based on previous words
\begin{algorithmic}[1]
	\State \texttt{$J$ $\leftarrow$ Number of Captions in Caption List}
	\State \texttt{$numEpoch$ $\leftarrow$ Number of epochs}
	\State \texttt{$batchSize$ $\leftarrow$ Number of batch size}
	\For{\texttt{$j$=1,...,$J$}}
	\State \texttt{Convert all words to lowercase in $\textbf{c}_j$}
	\State \texttt{Remove all punctuation in $\textbf{c}_j$}
	\State \texttt{Remove all words that are one character in length in $\textbf{c}_j$}
	\State \texttt{Remove all words with numbers in $\textbf{c}_j$}
	\EndFor	
	
	\State \texttt{represent $\textbf{C}$ with Word2Vec}

	\For{\texttt{$index$=1,...,$numEpoch$}}
	\For{\texttt{$indexBatchSize$=1,...,$batchSize$}}
	\State \texttt{1. Sample a mini batch of audio features $\textbf{x}$ }
	\State \texttt{2. Compute $p_\Theta(w_t|w_1,...,w_{t-1},\textbf{x}_j$)}
	\State \texttt{3. Update $\Theta$ by taking loss function on mini-batch loss according to the predicted partial caption. }
	\EndFor	
	\EndFor		
		
\end{algorithmic}
\end{algorithm}
\section{RESULTS}

In this section, we conduct our experiments on two publicly available audio captioning datasets, namely AudioCaps\cite{kim-etal-2019-audiocaps} and Clotho \cite{Drossos_2020}. To compare our results with existing methods, BLEU \cite{Papineni02bleu:a}, METEOR \cite{banerjee-lavie-2005-meteor}, CIDEr \cite{DBLP:journals/corr/VedantamZP14a}, and ROUGE$L$ \cite{lin-2004-rouge} metrics are used for the evaluations.

\subsection{Datasets}

AudioCaps is a large-scale dataset from AudioSet\cite{7952261}. It contains 46K 10-second video clips. For our experiments, we first extract audio files from the videos and build dataset splits for development, validation, and test splits, respectively. We have 45080 audio clips for development split, 487 audio clips for validation split, and 870 audio clips for test split. In the development split, every audio clip has one caption, but in the other splits, there are five captions for each clip. The word vocabulary size is 4364.

In the Clotho dataset, only development and evaluation parts are published. The development and the evaluation sets of the dataset contain 2893 and 1043 audio clips, respectively. Both of the sets have 5 captions for each audio clip. The length of the audio clips is 15 to 30 seconds in duration and captions are 8 to 20 words. We used the data splits as in \cite{Drossos_2020}. We use evaluation split for testing purposes. The word vocabulary size is 4366.

We use each audio clip five times with one assigned caption from the caption-list based on the best practice in \cite{Drossos_2020} for the Clotho dataset. For instance, let $a_i$ is an individual audio clip with captions  $S={[s_1,s_2,..,s_5]}$, then we use this audio clip instance as 5 separate instances: $<a_i,s_1>, <a_i,s_2>, .., <a_i,s_5>$ in the training.  We conduct a similar method for using five captions for validation split on the AudioCaps. To find the start and end of the sequences of the captions, we add special $<sos>$ and $<eos>$ in the beginning and end of the captions in both of the datasets.

\subsection{Training Details}

Our model has approximately 2,500,000 parameters. Adam optimizer and LeakyRelu activation function are used in the training. Batch-size is set to 64. We use a dropout rate of 0.5 for input connections. Batch-size and dropout rate are selected experimentally. Batch normalization \cite{DBLP:journals/corr/IoffeS15} is used after each BiGRU and GRU layer in the encoding and decoding phases. The loss function is categorical-cross entropy since it is widely used in the literature \cite{tanti-etal-2017-role}. It is given by 

\begin{equation}
\begin{aligned}
L(\Theta) =- \sum_{t=1}^{T} \log p_\Theta(w_t|w_1,...,w_{t-1})
\end{aligned}
\end{equation} 
where $w_t$ is the target word based on previous words.

To prevent gradient vanishing problem, LeakyReLU activation function is chosen empirically:

\begin{equation}
\begin{aligned}
LeakyReLU(x) =
\begin{cases}
x & \text{x$>$0}\\
\alpha & \text{x$\leq$0}\\
\end{cases} 
\end{aligned}
\end{equation}
where $\alpha$ is chosen 0.3 in this study which is the default value of LeakyReLU in Keras \cite{keras}. It uses a small gradient when the cell is not active.

The final hyperparameters such as the batch-size, dropout rate, and activation functions used in the study are chosen based on minimum validation loss in our several experiments. We implemented the system using Keras framework and run on a computer with GPU GTX1660Ti in a system Linux Ubuntu 18.04 and Python 3.6. The model is run for 50 epochs. In the experiments, 1 epoch with log-Mel energy features takes approximately 4 hours whereas 1 epoch with the PANNs features take only 5 minutes according to the given configurations. The minimum validation error is obtained in the $30^{th}$ epoch for the PANNs model given in Figure 5.
\begin{figure}[t]
	\centering
	\includegraphics[width=0.7\linewidth]{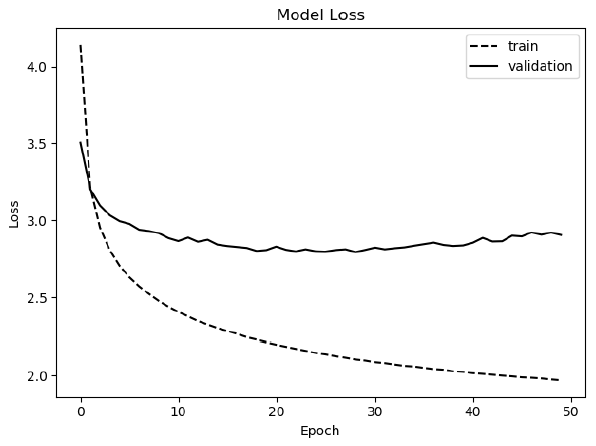}
	\vspace*{8pt}
	\caption{The train-validation loss of the model}
\end{figure}

\subsection{Evaluation}

We perform our evaluations on two public performance datasets AudioCaps and Clotho and compare our performance with the literature. We evaluate our experiments with widely used metrics in machine translation tasks: BLEU, METEOR, CIDEr, and ROUGE$L$.

We compare our results with the previous studies that we have introduced in detail in the Introduction section. 
 
 The metric BLEU${_n}$ calculates the precision for n-grams. To calculate precision, the matching words in the actual sentence and the predicted sentence is calculated. BLEU does not consider the context of the word in the sentence. The metric range is between [0,1]. If the actual sentence and the predicted sentence are totally the same, then the score is 1. BLEU-1 (B-1) represents 1-gram, whereas BLEU-4 (B-4) represents 4-grams. METEOR calculates recall and precision together and takes a harmonic mean score. It creates an alignment between actual and predicted sentences and makes mapping between them. CIDEr also uses n-gram model and it calculates cosine-similarity between the actual and predicted sentences. It also considers the Term Frequency Inverse-Document Frequency. ROUGE$_L$ calculates Longest Common Subsequences which considers the sequence of the words in the actual and predicted sentences.

 \subsection{Results}
 
 Our experimental results are presented in Table 1 and Table 2. Figure 6 and Figure 7 show our proposed methods outperform the state-of-the-art. 
 
 The results show that our proposed model RNN-GRU-EncDec with the log Mel features on the Clotho dataset has better results than the literature. The proposed model with the VGGish and the PANNs embeddings provides better results also it's training time is less than log Mel features. This is an expected result since the VGGish and PANNs are pretrained models. Also, training on the log Mel features consumes much more time. PANNs provide best results and training performance in terms of time and memory usage.
 
 Similarly, PANNs show best results on the AudioCaps dataset than the VGGish embeddings and log-Mel features. The RNN-GRU-EncDec architecture with the log Mel energies has lower performance than the previous study \cite{kim-etal-2019-audiocaps} on AudioCaps dataset since previous models on the AudioCaps dataset use pretrained VGGish embeddings. The RNN-GRU-EncDec architecture with the VGGish embeddings has comparable performance with the studies in the literature. Some of the metrics have lower values since the previous studies use semantic information in their model. When we add subject-verb embedding to our model as semantic information, our model outperforms the state-of-the-art.
 
  \begin{table*}[t]	
	\caption{Performance comparison of the proposed method with the Clotho Dataset. RNN-GRU-EncDec is the proposed encoder-decoder based architecture for sequence modeling. (SVE: Subject-Verb Embeddings)}
	\begin{center}
		\resizebox{\textwidth}{!}{
			\begin{tabular}{ l|l|c|c|c|c|c|c}
				\hline
				\multirow{2}*{\bfseries \hspace{2.5cm} Method} & \multicolumn{7} {c}{\bfseries Metric} \\
				\cline{2-8}
				& \textbf{B-1}& \textbf{B-2}& \textbf{B-3} & \textbf{B-4} & \textbf{CIDEr} & \textbf{METEOR} & \textbf{ROUGE$_L$}\\
				
				\hline
				\textbf{Clotho \cite{Drossos_2020}}    &       0.42     &   0.14 &  0.06      &    0.02   &    0.10 &    0.09 &    0.27\\
				
				\textbf{Temporal sub-sampling (M=16) \cite{nguyen2020temporal}}    &       0.43     &   0.15 &  0.06      &    0.02   &    0.09 &    0.09 &    0.27\\
				
				\textbf{CWR-WL-CAPS \cite{akr2020multitask}}    &       0.41     &   0.16 &  0.07      &    0.03   &    0.11 &    0.09 &    0.28\\
				
				\textbf{Proposed RNN-GRU-EncDec + Log Mel Energy \cite{eren2020audio}}   &       0.45    &  0.21 &   0.16      &    0.08  &    0.11 &    0.17 &    0.34 \\
				
				\textbf{Proposed RNN-GRU-EncDec + VGGish \cite{eren2020audio}}   &       0.51    &  0.28 &   0.22      &    0.12  &    0.18 &    0.19 &    0.40 \\
				
				\textbf{Proposed RNN-GRU-EncDec + PANNs}    &       0.57    &  0.34 &   0.25      &     0.14  &    0.28 &    0.21 &    0.44 \\
				
				\textbf{Proposed RNN-GRU-EncDec + PANNs + SVE}    &       \textbf{0.59}    &  \textbf{0.35} &   \textbf{0.26}      &     \textbf{0.14}  &    \textbf{0.28} &    \textbf{0.22} &    \textbf{0.45} \\
				\hline
			\end{tabular}
		}
	\end{center}
	\label{table1}
\end{table*}

\begin{table*}[t]
	
	\caption{Performance comparison of the different methods with the AudioCaps dataset. RNN-GRU-EncDec is the proposed encoder-decoder based architecture for sequence modeling. (SVE: Subject-Verb Embeddings)}
	\begin{center}
		\resizebox{\textwidth}{!}{
			\begin{tabular}{ l|l|c|c|c|c|c|c}
				\hline
				\multirow{2}*{\bfseries \hspace{2.5cm} Method} & \multicolumn{7} {c}{\bfseries Metric} \\
				\cline{2-8}
				& \textbf{B-1}& \textbf{B-2}& \textbf{B-3} & \textbf{B-4} & \textbf{CIDEr} & \textbf{METEOR} & \textbf{ROUGE$_L$}\\
				
				\hline
				\textbf{TempAtt-VGGish (C3)-LSTM \cite{kim-etal-2019-audiocaps}}    &       0.612     &   0.441 &  0.303      &    0.209   &    0.523 &    0.190 &    0.437\\
				
				\textbf{TopDown-VGGish (FC2,C4)-LSTM \cite{kim-etal-2019-audiocaps}}    &       0.629     &   0.451 &  0.315      &    0.214   &    0.577 &    0.199 &    0.448\\
				
				\textbf{TopDown-AlignedAtt (1NN) \cite{kim-etal-2019-audiocaps}}    &       0.614     &   0.446 &  0.317      &    0.219   &    0.593 &    0.203 &    0.450\\
				
				\textbf{Proposed RNN-GRU-EncDec + Log Mel Energy}    &       0.566    &  0.343 &   0.258      &     0.148  &    0.275 &    0.225 &    0.482 \\
				
				\textbf{Proposed RNN-GRU-EncDec + VGGish}    &       0.604    &  0.380 &   0.286      &     0.168  &    0.412 &    0.241 &    0.512 \\
				
				\textbf{Proposed RNN-GRU-EncDec + PANNs}    &       0.710    &  0.491 &   0.375      &     0.231  &    0.730 &    0.271 &    0.579 \\
				
				\textbf{Proposed RNN-GRU-EncDec + PANNs + SVE }    &       \textbf{0.711}    &  \textbf{0.493} &   \textbf{0.376}     &     \textbf{0.232}  &    \textbf{0.750} &    \textbf{0.287} &    \textbf{0.587} \\
				\hline
			\end{tabular}
		}
	\end{center}
	\label{table2}
\end{table*}

\begin{figure*}
	\centering
	\includegraphics[width=0.8\linewidth]{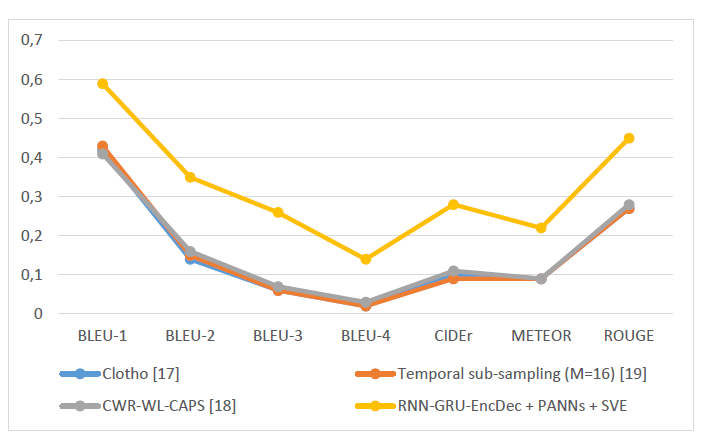}
	\caption{Performance comparison with existing methods on the Clotho dataset.}
\end{figure*}

 \begin{figure*}
	\centering
	\includegraphics[width=0.8\linewidth]{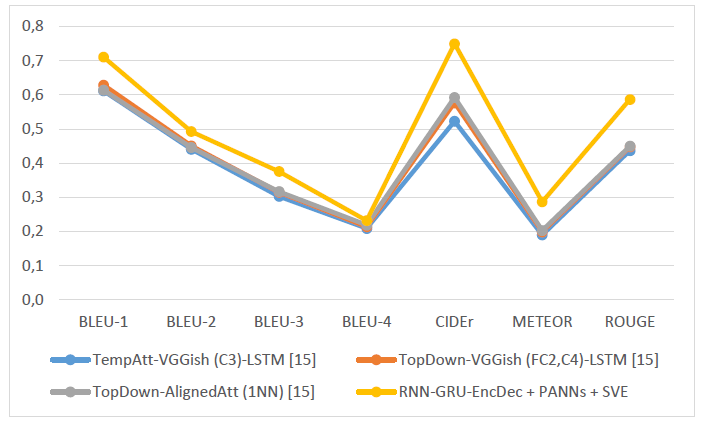}
	\caption{Performance comparison with existing methods on the AudioCaps dataset.}
\end{figure*}

Since the number of training data on the AudioCaps dataset is much more than the number of training data on the Clotho dataset, the AudioCaps dataset gives higher scores for all of our proposed models.

The inclusion of subject-verb embeddings yields better results on both of the datasets. Our results show that the inclusion of SVE improves the results on the AudioCaps dataset and provides more improvement on the Clotho dataset. The reason is that the Clotho dataset includes 5 sentences (labels) for each audio clip whereas the AudioCaps dataset has only one sentence (label) for each audio clip in the dataset. Though the improvement is minor on the AudioCaps dataset, our preliminary results show us the SVE can improve the results when multiple labels are used for each clip. We believe that using the SVEs may enhance success, especially when considered the subjectivity of the multi-labels of audio clips.  
 
 The predicted sentences show that our model can generally predict the main content of the audio clip. For instance, our model predicts ``{\textit{People are talking and laughing}'' whereas the ground truth is ``{\textit{People are talking and laughing with loud person near the end}''. It predicts the sentence in the correct order but shorter than the ground truth. 
 
 In our proposed model, similar concepts are also predicted. To illustrate, our model predicts ``{\textit{Rain is falling heavily and thunder is booming}'' while the ground truth is ``{\textit{Passing windstorm outside and something is striking against another harder object}''. Actually, they are similar concepts but according to BLEU, it is not assessed as a successful instance because the metric is based on calculating precision on exactly the same words. As another example, our model predicts the caption as ``{\textit{Bicycle is coasting down road slowly}'' whereas the ground truth is ``{\textit{The engine of vehicle is driving down the road}''. In this example, our model does not differentiate the bicycle and engine sounds. Some other predicted captions are given below to show our model’s performance for the Clotho and AudioCaps datasets.\\

 Clotho Dataset Examples:
 
\textbf{Actual-1}: {\fontfamily{qcr}\selectfont Blowing horn is followed by the siren from an emergency vehicle then the vehicle passes.}
 
\textbf{Actual-2}: {\fontfamily{qcr} \selectfont Police siren warns in four short bursts and then wails loudly as people are talking.} 
 
\textbf{Actual-3}: {\fontfamily{qcr} \selectfont Siren beeps many times then begins to wail constantly as it moves into the distance.} 
 
\textbf{Actual-4}: {\fontfamily{qcr} \selectfont Siren beeps several times then wails constantly as it moves into the distance. }
 
\textbf{Actual-5}: {\fontfamily{qcr} \selectfont Siren of car started blaring and the car drove off} 
 
 \textbf{Prediction}: {\fontfamily{qcr} \selectfont Siren is being played while people are talking in the background.} 
  \\
 
 \textbf{Actual-1}: {\fontfamily{qcr} \selectfont The footsteps of person are echoing as they are walking inside .} 
 
\textbf{Actual-2}: {\fontfamily{qcr} \selectfont Heavy footsteps resound in quiet open space.} 
 
\textbf{Actual-3}: {\fontfamily{qcr} \selectfont The quiet of place is disturbed by thudding footsteps.}
 
\textbf{Actual-4}: {\fontfamily{qcr} \selectfont The woman in high heels stomps across the stage before rustling papers.} 

\textbf{Actual-5}: {\fontfamily{qcr} \selectfont Person is walking inside with an echo footsteps.}
 
\textbf{Prediction}: {\fontfamily{qcr} \selectfont Someone is walking on the floor with the boots and echoing.}\\

AudioCaps Dataset Examples:

\textbf{Actual-1}: {\fontfamily{qcr} \selectfont Group of men speaking as cannons fire while rain falls and water splashes followed by thunder roaring.} 
 
\textbf{Actual-2}: {\fontfamily{qcr} \selectfont Man speaks then sudden explosion which is followed by smaller explosions and thunder.} 
 
\textbf{Actual-3}: {\fontfamily{qcr} \selectfont Male yelling and multiple gunshots.}
 
\textbf{Actual-4}: {\fontfamily{qcr} \selectfont Gunfire is ongoing and water is splashing adult males are shouting in the background and an adult male speaks in the foreground.} 
 
\textbf{Actual-5}: {\fontfamily{qcr} \selectfont Loud gunshots and explosions with men speaking water splashing wind blowing and thunder roaring.}
 
\textbf{Prediction}: {\fontfamily{qcr} \selectfont Man speaks followed by loud explosion and then man talking.}

\textbf{Actual-1}: {\fontfamily{qcr} \selectfont Rustling pigeons coo.} 

\textbf{Actual-2}: {\fontfamily{qcr} \selectfont Birds cooing and rustling.} 

\textbf{Actual-3}: {\fontfamily{qcr} \selectfont Pigeons coo and rustle.}

\textbf{Actual-4}: {\fontfamily{qcr} \selectfont Group of pigeons cooing. } 

\textbf{Actual-5}: {\fontfamily{qcr} \selectfont Pigeons are making grunting sounds and snapping beaks.}

\textbf{Prediction}: {\fontfamily{qcr} \selectfont Pigeons coo and flap wings.
}
 \section{CONCLUSION}
 
 In this paper, we present a novel encoder-decoder model, namely RNN-GRU-EncDec that combines text and audio features to predict audio captions using semantic and audio embeddings. We use the VGGish and PANNs audio embeddings to provide a smaller feature dimension compared to the raw audio features such as the log Mel band energies while preserving the performance. The subject-verb embeddings are used to show the semantic information contribution to audio captioning task. The results show that semantic information can improve audio captioning performance and audio embeddings bring us better training performance.
 
 The predicted captions show that our model is able to predict audio captions. We observe that, the generated captions are more general and shorter than the ground truths. Also, the proposed model does not very good at differentiating perceptually similar sounds such as ``bus'' and ``engine'' sounds. It can explicitly be stated that we can obtain better results if we have a larger dataset and train it for more on the powerful GPUs. Additionally, improving the language model and adding semantic information may increase the performance.
 
 According to these results, our future research direction is to strive for improving language modeling and to use data augmentation techniques in an attempt of enhancing the performance of our model. Getting better results on audio captioning can yield improvement in audio analysis. Additionally, multimodal models can be researched to improve the performance of video applications such as video captioning, video retrieval, and surveillance systems which are mainly composed of audio and video analysis.

\bibliographystyle{ws-ijsc}
\bibliography{sample}

\end{document}